\documentclass{ws-procs975x65}

\newcommand{\be}{\begin{equation}}
\newcommand{\ee}{\end{equation}}
\newcommand{\ba}{\begin{eqnarray}}
\newcommand{\ea}{\end{eqnarray}}

\begin{document}

\title{Flat space
modified particle dynamics induced by   Loop Quantum Gravity }

\author{LUIS F. URRUTIA\footnote{\uppercase{W}ork partially
supported by  the grants \uppercase{CONACYT}-40745-\uppercase{F}
and \uppercase{DGAPA-UNAM-IN}11700.}}

\address{Departamento de F\'\i sica de Altas Energ\'\i as \\
Instituto de Ciencias Nucleares\\
Universidad Nacional Aut\'onoma de M\'exico\\
Apartado Postal 70-543\\
04510 M\'exico, D.F. \\
E-mail: urrutia@nuclecu.unam.mx}

\maketitle

\abstracts{Starting from an heuristic approach to the
semiclassical limit in loop quantum gravity, the construction of
effective Hamiltonians describing Planck length corrections to the
propagation of photons and spin $1/2$ fermions, leading to
modified  energy-momentum relations, is summarized. Assuming the
existence of a privileged reference frame, we also review the
determination of stringent bounds upon the parameters labelling
such corrections, based upon already existing experimental data,
which are found to be from five to nine orders of magnitude below
the expected values.}

\section{Introduction}

\label{sec:intro}

The possibility of probing quantum gravity effects through minute
but detectable  modifications to standard particle dynamics has
sparked numerous  investigations regarding the derivation of such
corrections from a fundamental theory, together with detailed
studies of the signatures  identifying them.

A seminal proposal in this direction, leading to the research
field now called Quantum Gravity Phenomenology, was the work of
Amelino-Camelia et. al.,\cite{ACELLIS} suggesting that quantum
gravity effects would modify the standard photon energy-momentum
relation (c=1) in the form
 \be \label{DR} {\bf p}^2=E^2\left(1+
\xi\left(\frac{E}{E_{QG}}\right)+O\left(\left(\frac{E}{E_{QG}}\right)
\right)^2\right), \ee where $E_{QG}$ is a quantum gravity scale
expected to be  of the order of the Planck mass and $\xi$ is a
parameter of order one. Such modified dispersion relations have
been further generalized to massive particles.

Even though the above modifications are highly suppressed by the
quantum gravity scale, the possibility to amplify them by
observing time-delays of high energy photons detected in gamma-ray
burst originating  at cosmological distances has been
proposed.\cite{ACELLIS} Subsequently, additional astrophysical
phenomena (like ultra high energy cosmic rays (UHECR), neutrinos,
and radiation from  nearby BL Lac objects, for example), together
with atomic physics experiments have been studied in order to test
the proposed corrections. Many of these observations were shown to
impose stringent bounds upon the dimensionless parameters
labelling such modifications for photons and spin $1/2$
fermions.\cite{ANALYSIS,AP2002,SUDVUC,SUDVUC1}

An immediate consequence of the modified dispersion relation
(\ref{DR}) is a non-universal energy-dependent photon velocity,
which implies a violation of standard Lorentz covariance. This
point of view makes direct contact with a large body of research
related to the experimental determination of bounds upon the
parameters describing Lorentz invariance violation (LIV), which
started around 1960.\cite{VB,LIVMEETINGS} An extension of the
standard model, incorporating all possible LIV terms compatible
with the known high energy interactions has been
developed,\cite{KOSTELECKYMOD} which allows to correlate the
diverse experimental information. The model  has recently been
extended to include gravity.\cite{KOSTELECKYGRAV}

Different approaches have been followed to reproduce the proposed
dynamical modifications from a fundamental theory. Among them we
find  those arising from loop quantum gravity (LQG),
\cite{GP,URRUFOT,URRUFERM} and string
theory.\cite{STRING,ELLISFOT,ELLISNU}

A common feature of most theories describing quantum gravity is
the granular structure of space at distances of the order of the
Planck length ($\ell_P$), as opposed to the  continuum description
prevailing at large distances. In particular, one of the most
striking predictions of LQG is that the eigenvalues of the area
and volume operators are quantized in the corresponding units of
Planck length,\cite{ROVSMOL} thus invalidating the continuum
description of space at very short distances. In analogy with wave
propagation in a crystal, for example, one indeed expects that
such granularity would induce dynamical corrections with respect
to the propagation in the continuum.

Some points of view that have been adopted in the literature in
relation to the effects of this granularity are :

 (i) Dynamical corrections do arise, which signal the breaking of standard
Lorentz covariance. In particular this implies the existence of a
privileged reference frame (a return to the idea of the ether)
which has been usually identified with the system where the cosmic
microwave background radiation (CMB) looks isotropic. This point
of view makes direct contact with the above mentioned experimental
searches for the breaking of Lorentz invariance.

(ii) Dynamical corrections do arise, but a new relativity
principle is introduced by deforming or extending  the standard
Lorentz transformations, so that no privileged reference frame
appears. These proposals go under the name of Double Special
Relativity (DSR) and basically include some maximum energy
(Planck's energy for example) as an additional invariant in the
modified Lorentz transformations, besides the standard speed of
light as the maximum attainable velocity.\cite{DSR}

(iii) Standard Lorentz covariance is preserved in spite of the granular
character of space at very short distances.\cite{ROVSOR}

In the context of LQG, the  effective classical Hamiltonian for
each kind of particle can be  constructed as  the expectation
value of the corresponding quantum gravity operator in an adequate
semiclassical state of the Hilbert space describing the prescribed
classical matter field together with a continuum metric at large
distances (flat space, for example), while retaining the quantum
discreteness properties at short distances.

The first ingredient of this approach has been already developed
in Thiemann's proposal for the construction of  regularized
versions of the required quantum Hamiltonians.\cite{ThiemannR} The
construction of exact semiclassical states has proved to be more
elusive and still constitutes  an open problem, in spite of the
progress already made. In fact, these more elaborate calculations
provide additional support to the existence of Planck scale
modifications to the dynamics.\cite{THIEMANN2} In this way, the
LQG inspired calculations of the effective Hamiltonians to be
reviewed in this work are based upon some reasonable and general
assumptions regarding the behavior of the required operator
expectation values under the {\it would be} semiclassical states
defined in the kinematical Hilbert space of LQG.

The paper is organized as follows. Section \ref{sec:genidea}
provides a brief review of the main ideas and methods employed by
Alfaro, Morales-T\'ecotl and Urrutia to estimate such corrections
in the case of photons and spin $1/2$ fermions. The corresponding
results are summarized in Section
\ref{sec:results}.\cite{URRUFOT,URRUFERM}  Assuming the existence
of the privileged CMB frame, stringent bounds upon the LIV
parameters in the fermionic case arising from clock-comparison
experiments in atomic physics are reviewed in Section
\ref{sec:bounds}.\cite{SUDVUC} Finally, some closing comments are
given in Section \ref{sec:final}.

\section{Corrections to standard particle dynamics in flat space }
\label{sec:genidea} Central to the approach of Alfaro,
Morales-T\'ecotl and Urrutia \cite{URRUFOT,URRUFERM} is Thiemann's
regularization of the LQG Hamiltonians ${\hat
H}_{\Gamma}$.\cite{ThiemannR} This is based upon a triangulation
of space, adapted to the corresponding graphs $\Gamma$ which
define a given state in the loop representation.

The quantum Hamiltonians are defined via the holonomies and fluxes
of the quantized connections and canonically conjugated momenta,
respectively, around and through the faces of the tetrahedra
defining the triangulation. The regularization is provided by the
volume operator, with discrete eigenvalues arising only from the
vertices which are common to the graph and the triangulation.
Those vertices are the only ones that contribute to the action of
the operator in the wave function.

Here we take an heuristic point of view, starting from the exact
operator version of LQG and defining its action upon the {\it
would be}  semiclassical states through some plausible
requirements.

We think of the semiclassical configuration describing a
particular matter or gauge field operator ${\hat F}$ plus gravity,
as given by an ensemble of graphs $\Gamma$, each occurring with
probability $P(\Gamma)$. To each of such graphs we associate a
wave function $|\Gamma, {\mathcal L}, F \rangle$ which is peaked
with respect to a classical  field configuration $F$, together
with a flat gravitational metric and a zero value for the
gravitational connection at large distances. In other words, the
contribution of the gravitational operators inside the expectation
value is estimated as
\begin{eqnarray}
\langle\Gamma, {{\mathcal L}}, F|\, ...{\hat q}_{ab}...\,|\Gamma,
{{\mathcal L}}, F \rangle&=& \delta_{ab} + O\left(\frac{\ell_P}{
{\mathcal L}}\right),\nonumber\\
\langle\Gamma, { {\mathcal L}}, F |\, ...{\hat
A}_{ia}...\,|\Gamma, {{\mathcal L}}, F \rangle&=& 0\, + \frac{1}{
{\mathcal L}}\, \left(\frac{\ell_P}{{ {\mathcal
L}}}\right)^\Upsilon. \label{EXPV}
\end{eqnarray}
The parameter $\Upsilon \geq 0$ is a real number.
 Also we associate the effective Hamiltonian $H_{\Gamma}=
 \langle \Gamma,{\mathcal L},F\,|{\hat H}_{\Gamma}
 |\, \Gamma,{\mathcal L},F\rangle$  to each graph.
 The coarse graining scale ${ {\mathcal L}}>> \ell_P \, $ of the
 wave function is such that the continuous
 flat metric approximation is appropriate for
 distances much larger that ${\mathcal L}$,
 while the granular structure of space becomes
 relevant when probing
 distances smaller that ${\mathcal L}$. In this way,
 space is constructed by
 adding boxes of size ${\mathcal L}^3$, which center
 represents a given point $\bf x$
in the  continuum and which contain a large number of vertices of
the adapted triangulation, together with the corresponding
tetrahedra.

 The  field $\hat{F}$, characterized by a De Broglie wave length
 $\lambda$,
 is considered a slowly varying function within each box ( $\lambda >
 {\mathcal L}$) and  is expanded in power
 series of the segments of the tetrahedra
 having a common vertex with the graph.
 The contribution of each  of these segments to
 the expectation value is estimated by $\ell_P$.
Also, under the expectation value, the  contribution of $\hat{F}$
is given
 by the value of the classical field and its derivatives at the center
 of the box. Gravitational
 variables are rapidly varying inside the box.

The total effective Hamiltonian is defined as an  average over the
graphs $\Gamma$ which define the semiclassical limit: ${\rm
H}=\sum_{\Gamma} P(\Gamma)\, {\rm H}_\Gamma $. This effectively
amounts to average the expectation values of the gravitational
variables in each box. We construct such averages in terms of  the
most general combinations of flat space tensors $\delta_{ab},
\epsilon_{abc}, \dots $ which saturate the tensor structure of the
classical fields together with their derivatives in each box. In
this way we are imposing rotational invariance on our final
effective Hamiltonian.

Next we make some general comments regarding the above procedure:

(i) our calculation has been performed in a fixed reference frame
and  leaves undetermined an overall numerical dimensionless
coefficient in each of the calculated corrections. If these
coefficients are non-zero, one would expect them to be or order
one, meaning that the physics at the Planck scale has been
correctly taken into account. The results can be viewed as an
expansion in terms of the classical fields and their derivatives,
combined with an explicitly factored out dependence upon the two
scales $\ell_P$ and ${\mathcal L}$.

(ii) The non-zero corrections obtained in this way have been
usually interpreted as signaling a preferred reference frame
together with a violation of the standard active (particle)
Lorentz transformations. The advent of DSR opens up the
possibility to study whether or not such modified actions can be
embedded in a related framework, thus recovering a modified
relativity principle and eliminating the appearance  of a
privileged reference frame. Also, there  is the  possibility that
a full-fledged calculation of the correction coefficients would
produce a null result, thus enforcing standard Lorentz covariance.
\section{The results}
\label{sec:results} In this section we summarize the calculated
effective Hamiltonians together with the corresponding modified
dispersion relations, for the cases of photons and two-component
spin 1/2 particles.
\subsection{Photons}
\label{ssection: photons} The resulting effective Hamiltonian
is\cite{URRUFOT} \ba \label{HEMFIN} {\rm H}^{EM}&=&
\frac{1}{Q^2}\int d^3{\vec x} \left[\left(1+ \theta_7
\,\left(\frac{\ell_P}{{ {\mathcal L}}}\right)^{2+2\Upsilon}
\right)\frac{1}{2}\left(\frac{}{} \underline{{\vec B}}^2 +
\underline{{\vec E}}^2\right) + \theta_2\,\ell_P^2\,{\underline
E}^a
\partial_a \partial_b {\underline E}^b+
\right. \nonumber \\
&&\left.+ \theta_3 \, \ell_P^2 \, \left( \frac{}{}\underline{B}^a
\,\nabla^2 \underline{B}_a + \underline{E}^a \,\nabla^2
\underline{E}_a\right)+\theta_8 \ell_P \left( \frac{}{}
\underline{\vec B}\cdot(\nabla \times\underline{\vec B})+
\underline{\vec E}\cdot(\nabla \times\underline{\vec E}) \right)+
\ \right.  \nonumber \\
&&\left.  + \theta_4\, { {\mathcal L}}^2 \, \ell_P^2 \,
\left(\frac{{ {\mathcal L}}}{\ell_P} \right)^{2 \Upsilon}\,
\left(\frac{}{}\underline{{\vec B}}^2\right)^2 +\dots \right], \ea
up to order $\ell_P^2$. Here $\theta_i$ label the parameters left
undetermined by our procedure and $Q$ is the gauge
(electromagnetic) coupling in Thiemann's notation.

The corresponding dispersion relation is
\begin{eqnarray}
\omega_{\pm}= k\left(1+\theta_7\left(\frac{\ell_P}{{ {\mathcal L}}
}\right)^{2+2\Upsilon}-2\,\theta_3\,(k\ell_P)^2\pm
2\theta_8\,(k\ell_P ) \right).
\end{eqnarray}
The $\pm$ signs correspond to the two polarizations of the photon.
The speed of the photon is given by $ v_{\pm}(k, { {\mathcal L}})=
{\partial \omega_{\pm}(k, { {\mathcal L}})}/{\partial k}$.
Choosing ${\mathcal L}=\lambda=1/k$, we recover the dominant
helicity dependent correction found already in the seminal work of
Gambini and Pullin.\cite{GP} As far as the $\Upsilon$ dependent
terms we have either a quadratic ($\Upsilon=0$) or a quartic
($\Upsilon=1$) correction. The only possibility to have a first
order helicity independent correction amounts to set
$\Upsilon=-1/2$ which corresponds to that of Ellis et. al..
\cite{ELLISFOT} However, we do not have an interpretation for such
a value of $\Upsilon$.

First steps towards the generalization of the Hamiltonian
(\ref{HEMFIN}) to the Yang-Mills case have been
taken.\cite{URRUYM} In this work the holonomy of a non-abelian
connection in an arbitrary triangular path appropriate to a face
of the tetrahedra defining the triangulation  has been calculated
in powers of the corresponding segments of the triangulation, up
to fifth order. One expects the non-abelian results to be obtained
from those of the photon case just by changing ordinary
derivatives $\partial_a$ into covariant derivatives $D_a$.
Nevertheless, this procedure does not produce a unique answer when
dealing with more that one derivative, since $[D_a, D_b]\neq 0$.
Thus, to resolve the ambiguity of such guessing one has to perform
the complete calculation.

\subsection{Two-component spin 1/2 particles}

\label{ssection: fermions}

The effective Hamiltonian is\cite{URRUFERM}

\begin{eqnarray}  \label{EFFHF}
{\rm H}_{1/2} = \int d^3 x \left[ i \ \pi(\vec x) \tau^d\partial_d
\ {\hat A} \right. \xi({\vec x}) + c.c.  + \frac{i}{4\hbar}
\frac{1}{{ {\mathcal L}}} \ \pi({\vec x}) \,{\hat
C}\, \xi({\vec x})\quad \qquad  &&\nonumber\\
 + \frac{m}{2 \hbar } \xi^T({\vec
x})\ (i \sigma^2) \left( \alpha + 2\hbar\,\beta \ \tau^a
\partial_a \right)\xi({\vec x})
+\left. \frac{m}{2 \hbar} \pi^T({\vec x}) \left( \alpha  +
2\hbar\,\beta \  \tau^a \partial_a \right) (i \sigma^2) \pi({\vec
x}) \right],&&
\end{eqnarray}
where
\begin{eqnarray}\label{EFFHF1}
&&{\hat A}=\left(1 +
{ \kappa}_{1} \left(\frac{\ell_P}{{ {\mathcal L}}} \right)^{\Upsilon+1}+ { \kappa}%
_{2} \left(\frac{\ell_P}{{ {\mathcal L}}} \right)^{2\Upsilon+2} +
\frac{{ \kappa}_3}{2} \
\ell_P^2 \ \ \nabla^2 \right),\nonumber\\
&&{\hat C}=\hbar \ \left({ \kappa}_4 \left( \frac{\ell_P}{
{\mathcal L}}\right)^\Upsilon
+ { \kappa}_{5} \left(\frac{\ell_P}{%
{ {\mathcal L}}}\right)^{2\Upsilon+1}
+ { \kappa}_{6}\left(\frac{\ell_P}{{ {\mathcal L}}} \right)^{3\Upsilon+2} +\frac{%
{ \kappa}_{7}}{2} \left(\frac{\ell_P}{ {\mathcal
L}}\right)^{\Upsilon}\ \ell_P^2 \ \ \nabla^2\right),
\nonumber \\
&&  \alpha= \left(1 + { \kappa}_{8} \left(\frac{\ell_P}{{
{\mathcal L}}}\right)^{\Upsilon+1}\right),
 \qquad \beta=\frac{{%
 \kappa}_9}{2\hbar} \ell_P + \frac{\kappa_{11}}{2\hbar}\ell_P
 \left(\frac{\ell_P}{ {\mathcal L}} \right)^{\Upsilon+1}.
\end{eqnarray}
Here $\kappa_i$ are the undetermined coefficients,
$\tau_i=-(i/2)\sigma^i$ (where $\sigma^i$ are the standard Pauli
matrices), $\pi=i\xi^{*}$ and $m$ is the mass of the fermion.

The corresponding dispersion relation is
\begin{eqnarray}
E_\pm(p, { {\mathcal L}})&=&\left[ p+\frac{m^{2}}{2p}\pm \ell
_{P}\left( \frac{1}{2}m^{2}\kappa
_{9}\right) +\ell _{P}^{2}\left( -\frac{1}{2}\kappa _{3}p^{3}+\frac{1}{8}%
\left( 2\kappa _{3}+\kappa _{9}^{2}\right) m^{2}p\right) \right]\nonumber  \\
&+&\left( \frac{\ell _{P}}{{ {\mathcal L}}}\right) ^{\Upsilon+1
}\left[ \left( \kappa _{1}p-\frac{\Theta _{11}m^{2}}{4p}\right)
\pm \ell_P\,\left( -\kappa _{7}\frac{p^{2}}{4}+\Theta _{12}\frac{m^{2}%
}{16}\right) \right] \nonumber\\
&+&\left( \frac{\ell _{P}}{{ {\mathcal L}}}\right) ^{2\Upsilon
+2}\left( \kappa _{2}p-\frac{m^{2}}{64p}\Theta _{22}\right),
\label{CDR2}
\end{eqnarray}
where the new  coefficients $\Theta$ are linear combinations of
some  $\kappa$'s. The velocity of propagation is $ v_\pm(p, {
{\mathcal L}})= {\partial E_\pm(p, { {\mathcal L}})}/{\partial
p}$.

Alternative results based on a string theory inspired approach
have been reported in the literature.\cite{ELLISNU}

\subsection{The parameters ${{\mathcal L}}$ and $\Upsilon$ }

\label{ssection: param}

In order to produce numerical estimations of some of the effects
arising from the previously obtained modifications to flat space
dynamics, we must further fix the value of the scales $ {\mathcal
L}$ and $\Upsilon$.

Recall that ${\mathcal L}$ is a coarse graining scale indicating
the onset distance from where the non perturbative states in the
loop representation can be approximated by the classical flat
metric. The propagating particle  is characterized by energies
which probe distances of the order of the De Broglie wave length
$\lambda$. Just to be consistent with a description in terms of
classical continuous equations it is necessary to require that ${
{\mathcal L}}< \lambda$. Two distinguished cases arise: (i) the
mobile scale, where we take the marginal choice ${ {\mathcal L}}=
\lambda$  and (ii) the universal scale, which has been introduced
in the discussion of the GZK anomaly.\cite{AP2002} The
consideration of the different reactions involved produces a
preferred bound on ${ {\mathcal L}}: \, 4.6\times
10^{-8}GeV^{-1}\geq{ {\mathcal L}}\geq 8.3 \times
10^{-9}GeV^{-1}$. A recent study of the gravitational Cerenkov
effect together with neutrino oscillations \cite{LAMBIASE} yields
a universal scale estimation which is consistent with the former .

Bounds for $\Upsilon$ have been estimated  based on the
observation that atmospheric neutrino oscillations at average
energies of the order $10^{-2}-10^2 $ GeV are dominated by the
corresponding mass differences via the oscillation length $L_m$.
This means that additional contributions to the oscillation
length, in particular the quantum gravity correction $L_{QG}$,
should satisfy $L_{QG}> L_m$. This is used to set a lower bound
upon $\Upsilon.$ Within the proposed two different ways of
estimating the scale ${{\mathcal L}}$ of the process we obtain:
(i) $\Upsilon > 0.15$ when ${ {\mathcal L}}$ is considered as a
mobile scale and (ii) $\Upsilon > 1.2  $ when the scale ${
{\mathcal L}}$ takes the universal value ${{\mathcal L}}\approx
10^{-8}\, GeV^{-1}$.\cite{URRUFERM}

\section{Observational bounds from spin $1/2$ fermions using existing data}

\label{sec:bounds}

The previously found Hamiltonian was obtained under the assumption
of flat space isotropy and was assumed to  account for the fermion
dynamics in a preferred reference frame, identified  as the one in
which  the Cosmic Microwave Background looks isotropic. The earth
velocity $\mathbf w$ with respect to that frame has already been
determined to be $w/c\approx 1.23 \times 10^{-3}$ by COBE. Thus,
in  the earth reference frame one expects the appearance of
signals indicating minute violations of space isotropy encoded in
$\mathbf w$-dependent terms appearing in the transformed
Hamiltonian or Lagrangian.\cite{SUDVUC}  On the other hand, many
high precision experimental test of rotational symmetry, using
atomic and nuclear systems, have been already reported in the
literature.\cite{LIVMEETINGS} Amazingly enough such precision is
already adequate to set very stringent bounds on some of the
parameters arising from the quantum gravity corrections.

We have considered the case of non-relativistic Dirac particles
and obtained corrections which involve the coupling of the spin to
the CMB velocity together with a quadrupolar anisotropy of the
inertial mass.\cite{SUDVUC} The calculation was made with the
choices $\Upsilon=0$ and ${ {\mathcal L}}=1/M$, where $M$ is the
rest mass of the fermion. Keeping only terms linear in $\ell_P$,
the equation of motion arising from the two-component Hamiltonian
(\ref{EFFHF}) can be readily extended to the Dirac case as
\begin{equation}
\left( i\gamma ^{\mu }\partial _{\mu }\,+\Theta_1 m\ell
_{P}\;i\mathbf{\gamma}\cdot \nabla -\frac{K}{2}\gamma _{5}\gamma
^{0}-m\left( \alpha -i\Theta_2 \ell _{P}\,{{\Sigma}}\cdot \nabla
\right) \right) \Psi =0, \label{DIREQ}
\end{equation}
where we have used the representation in which $\gamma _{5}$ is
diagonal. The spin operator is\ $\Sigma ^{k}=(i/2)\epsilon
_{klm}\gamma ^{l}\gamma ^{m}$, $K=\Theta_4\,m^2\,\ell_P$ and
$\alpha=1+\Theta_3\,m\,\ell_P$. The normalization has been chosen
so that in the limit $(m\ell_P)\rightarrow 0$ we recover the
standard massive Dirac equation. The term $m\left(
1+{\Theta_3\;}m\ell _{P}\right) $ can be interpreted as a
renormalization of the mass whose physical value is taken to be
$M=m\left( 1+{\Theta_3\;}m\ell _{P}\right) $. After this
modifications we can write an effective Lagrangian describing the
time evolution as seen in the CMB frame. In order to obtain the
dynamics  in the laboratory frame we implement an observer Lorentz
transformation in the former Lagrangian and rewrite it  in a
covariant looking form by introducing explicitly the CMB frame's
four velocity $W^{\mu }=\gamma (1,\,{\mathbf{w}}/c)$. The result
is
\begin{eqnarray}
L_{D}&=&\frac{1}{2}i\bar{\Psi}\gamma ^{\mu }\partial _{\mu }\Psi -\frac{1}{2}M%
\bar{\Psi}\,\Psi +\frac{1}{2}i(\Theta_1 M\ell
_{P})\bar{\Psi}\gamma _{\mu }\left(
g^{\mu \nu }-W^{\mu }W^{\nu }\right) \partial _{\nu }\Psi \nonumber \\
&&+\frac{1}{4}(\Theta_2M\ell_{P})\bar{\Psi}\epsilon _{\mu \nu
\alpha \beta }W ^{\mu }\gamma ^{\nu }\gamma ^{\alpha }\partial
^{\beta }\Psi - \frac{1}{4}(\Theta_4 M\ell_P)M W_\mu
\bar{\Psi}\gamma_5\gamma^\mu\Psi +h.c. . \label{otra}
\end{eqnarray}
From the work of Kosteleck\'y and Lane\cite{Lane} we directly
obtain the non-relativistic limit of the Hamiltonian corresponding
to (\ref{otra}), up to first order in $\ell_P$ and up to order
${({\bf w})/c}^{2}$, which is
\begin{eqnarray}
\tilde{H}=  Mc^2(1+\Theta_1\, M\ell_P\,\left({\mathbf{w}}/{c}
\right) ^{2})+ \Theta_1M\ell _{P} %
\left[ \frac{{\bf w}\cdot {Q}_{P}\cdot {\bf w}}{Mc^2}\right] &&\nonumber\\
+ \left(1+2\,\Theta_1M\ell
_{P}\left(1+\frac{5}{6}\left({\mathbf{w}}/{c} \right)
^{2}\right)\right)\left(\frac{p^{2}}{2M}+g\,\mu
\,{\bf s}%
\cdot {\bf B}\right)&&\nonumber\\
 + \left(\Theta_2+\frac{1}{2}\Theta_4 \right)M\ell
_{P}\left[\left(2Mc^2
-\frac{2p^{2}}{3M}\right)\,{\bf s}%
\cdot \frac{\bf w}{c}+\frac{1}{M}\,{\bf s}\cdot {Q}_{P}\cdot
\frac{\bf w}{c}\right] ,&& \label{B6}
\end{eqnarray}
where ${ s^i}={ \sigma^i}/2$.

The above effective Hamiltonian has been used in the description
of the valence nucleon responsible for the transitions measured in
clock-comparison experiments using pairs of nuclei like
$({}^{21}Ne,{}^3He)$,\cite{Exper2} and
$({}^{129}Xe,{}^3He),$\cite{Exper3} for example. In (\ref{B6}) we
have not written the terms linear in the momentum since they
average to zero in the nuclear bound state situation. Here $g$ is
the standard gyromagnetic factor, and $Q_{P}$ is the momentum
quadrupole tensor with components
$Q_{Pij}=p_{i}p_{j}-1/3p^{2}\delta _{ij}$. The terms in the second
square bracket of the LHS of (\ref{B6}) represent a coupling of
the spin to the velocity of the privileged reference frame. The
first term inside the bracket has been measured with high accuracy
and an upper bound for the coefficient has been found. The second
term in the same bracket is a small anisotropy contribution and
can be neglected. Thus we find the correction \be \delta H_{S}=
\left(\Theta_2 +\frac{1}{2}\Theta_4\right) M\ell _{P} (2Mc^2)
\left[ 1 + O\left( \frac{p^{2}}{2M^2c^2}\right) \right]
\mathbf{s}\cdot \frac{\mathbf{w}}{c}. \ee

The first square bracket in the LHS of (\ref{B6})  represents an
anisotropy of the inertial mass and has been bounded in
Hughes-Drever like experiments. With the approximation
$Q_{P}=-5/3<p^{2}>Q/R^{2}$ for the momentum quadrupole moment,
with $Q$ being the electric quadrupole moment and $R$ the nuclear
radius, we obtain
\begin{equation}
\delta H_{Q}=-\Theta_1 M\ell _{P}\frac{5}{3}\left\langle \frac{p^{2}}{2M}%
\right\rangle \left( \frac{Q}{R^{2}}\right) \left(
\frac{w}{c}\right) ^{2}P_{2}(\cos \theta ), \label{QMM}
\end{equation}
for the quadrupole mass perturbation, where $\theta$ is the angle
between the quantization axis and $\mathbf{w}$. Using
$<p^{2}/2M>\sim 40$ MeV for the energy of a nucleon in the last
shell of a typical heavy nucleus, together with the experimental
bounds of references \cite{Exper2,Exper3} we find \cite{SUDVUC}
\begin{equation}
\mid \Theta_2+\frac{1}{2}\Theta_4\mid <2\times 10^{-9},\qquad \mid
\Theta_1\mid <3\times 10^{-5}. \label{Result2}
\end{equation}
The second bound in (\ref{Result2}) also imposes  stringent
constraints\cite{SUDVUC1} upon some string theory inspired models
that induce Planck scale corrections to field
propagation.\cite{ELLISNU}

 The above  bounds  on terms that were formerly expected to be
of order unity, already call into question the scenarios inspired
on the various approaches to quantum gravity suggesting the
existence of Lorentz violating Lagrangian corrections which are
linear in Planck's length. In relation to this point it is
interesting to observe that a very reasonable agreement with the
current AGASA  ultra high energy cosmic ray (UHECR) spectrum,
including the region beyond the GZK cutoff, has been recently
obtained by using dispersion relations of order higher than linear
in $\ell_P$, together with consideration of additional stringent
bounds arising from  the first estimations of the impact of nearby
BL Lac objects and UHECR data upon LQG parameters.\cite{AP2002}

\section{Final Comments}
\label{sec:final}

Since an exact construction of the semiclassical approximation in
LQG is still lacking, the heuristic  approach reviewed here
offers a framework to make some progress towards a final
understanding of the problem, together with its observational
implications. On the other hand, the stringent bounds found under
the assumption of the existence of a privileged frame already
forbid corrections to the dynamics which are linear in $\ell_P$,
within this scenario. From a purely phenomenological point of view
one could study the possibility to alleviate these constraints by
selecting an adequate parameter $\Upsilon$. Nevertheless a more
fundamental interpretation would still be lacking. Even though the
analysis of the modified dynamics in terms of a privileged
reference frame has been widely used in the literature, the
approach presented here is not necessarily bounded to the
existence such frame. In fact, the advent of DRS has provided
support to the coexistence between Planck-scale modified dynamics
and an extended relativity principle. The possibility to embed our
approach in a DSR-like framework needs to  be further explored,
having an eye on the implications that this requirement might
teach us regarding de structure of the much sought semiclassical
states, together with  identifying  new observational test that
the extended covariance will demand.

\end{document}